\documentclass[twocolumn,amsmath,amssymb,nofootinbib,eqsecnum,tighten]{revtex4}

\usepackage{graphicx}
\usepackage{dcolumn} 
\usepackage{bm,bbm}      
\usepackage{curves}
\usepackage{epic}
\usepackage{amsthm}
\usepackage[mathcal]{euscript}
\usepackage{epsf,times}

\def\=d{\, {\buildrel \rm def  \over =} \,}

\def\sqr#1#2{{\vcenter{\vbox{\hrule height.#2pt \hbox{\vrule width.#2pt height#1pt \kern#1pt \vrule width.#2pt}\hrule height.#2pt}}}}

\def\beq#1{\begin{equation} \label{#1}}
\def\eeq{\end{equation}}
\def\ben{\begin{equation*}}
\def\een{\end{equation*}}
\def\bequa{\begin{eqnarray}}
\def\eequa{\end{eqnarray}}
\def\re#1{(\ref{#1})}

\def\bf#1{\bm{#1}}

\def\path{figures}

\begin{document}

\title{Theory for Inelastic Neutron Scattering in Orthorhombic
High-T$^{\ }_c$ Superconductors}

\author{Andreas P. Schnyder$^{1}$, Dirk Manske$^2$, Christopher Mudry$^1$, and Manfred
Sigrist$^2$} \affiliation{$^1$ Condensed Matter Theory Group, 
Paul Scherrer Institute, CH-5232
Villigen PSI, Switzerland \\ $^2$Institut f\"ur Theoretische
Physik, ETH Z\"urich, H\"onggerberg, CH-8093 Z\"urich,
Switzerland}

\date{\today}

\begin{abstract}
Using a Fermi-liquid-based theory we calculate the in-plane
anisotropy of the spin susceptibility $\chi({\bf q},\omega)$ for
hole-doped high-$T^{\ }_c$ cuprates. Employing the two-dimensional
one-band Hubbard model and a generalized RPA-type theory we
consider anisotropic hopping matrix elements ($t^{\ }_x \neq t^{\ }_y$) and a
mixing of $d$- and $s$-wave symmetry of the superconducting order
parameter in order to describe orthorhombic superconductors. We
compare our calculations with available inelastic neutron
scattering data on untwinned
$\mbox{YBa}^{\ }_2\mbox{Cu}^{\ }_3\mbox{O}^{\ }_{6+x}$ and find good
agreement. Furthermore, we predict a strongly anisotropic 
in-plane dispersion of the resonance peak.
\end{abstract}

\maketitle


\section{Introduction}
The spin dynamics plays an essential role in high-$T^{\ }_c$ cuprates.
Superconductivity occurs very close to a Mott insulating state
supporting a strong long-range antiferromagnetic (AF) order. 
Inelastic neutron scattering (INS) demonstrates, 
through the so-called commensurate and incommensurate peaks,
the existence of magnetic collective phenomena in the superconducting 
state of hole-doped high-$T^{\ }_c$ cuprates
intimately tied to superconductivity.~%
\cite{bourges,keimer,sciencebourges,pailhes,hayden04,tranquada04,christensen04}

Several theoretical scenarios have proposed a mechanism
for superconductivity in the high-$T^{\ }_c$ cuprates attributed to magnetism.
It has been argued that superconducting quasiparticles
emerge from an exchange of AF spin fluctuations
between Fermi-like quasiparticles~%
\cite{scalapino,chubukov,dmbook}
or from a recombination in momentum space of holons and spinons 
in a spin-charge-separated normal state.%
\cite{Lee05}
In the stripe scenario,~%
\cite{Kivelson03}
strong electronic interactions
result in normal and superconducting states in which
spin and charge are separated in a predominantly one-dimensional
region, called stripes, of the CuO$^{\ }_2$ planes.

The incommensurate and commensurate peaks
seen in INS on 
$\mbox{La}^{\ }_{15/8}\mbox{Ba}^{\ }_{1/8}\mbox{Cu}\mbox{O}^{\ }_{4}$
in Ref.~\onlinecite{tranquada04}
are interpreted in terms of excitation spectra 
in a bond centered stripe state with quasi or long-range
magnetic order in the stripe picture of Refs.~\onlinecite{vojta,urig,seibold}. 
Using a Fermi-liquid-like theory for itinerant quasiparticles, 
it was argued in Refs.~%
\onlinecite{
bulut96,%
morr,%
dmiekhb01,%
brinckmann01,
yamase01,%
onufrieva,%
chubuk1,%
prelovsek,%
li02,%
schnyder04,%
ereminprl05%
}
that the incommensurate and resonance INS peaks in 
$\mbox{YBa}^{\ }_2\mbox{Cu}^{\ }_3\mbox{O}^{\ }_{6+x}$ (YBCO) 
or
Bi$^{\ }_2$Sr$^{\ }_2$CaCu$^{\ }_2$O$^{\ }_{8+x}$ (Bi2212)
are a fingerprint of a pure $d^{\ }_{x^2-y^2}$-wave symmetry
of the superconducting order parameter. 
In order to distinguish between the Fermi-liquid and
stripe pictures applied to YBCO, 
a detailed analysis of the spin excitations in \textit{untwinned} YBCO is helpful. 

Pure $d^{\ }_{x^2-y^2}$-pairing symmetry is only to be expected for
underlying lattices with tetragonal symmetry. 
Most of the cuprates are known to show orthorhombic distortions.
The high-$T^{\ }_c$ superconductor
YBCO
reveals a strong structural orthorhombic distortion 
as a function of doping.
For example, a 60 $\%$ anisotropy 
in the London penetration depth 
between the $a$ and $b$ directions in the two-dimensional
CuO$^{\ }_2$ planes was found
by Basov \textit{et al.}~\cite{basov95}
As YBCO is characterized by CuO-chains that 
are present only along the $b$ direction
and as these chains are believed
to act as charge reservoirs that fill up with 
increasing doping $x$, a density functional calculation
predicts a distorted Fermi surface (FS) in two dimensions.~\cite{lichti} 
This prediction of a two-dimensional anisotropic FS is
consistent with angle-resolved photoemission
spectroscopy (ARPES) studies by Lu \textit{et al.} \cite{lu01}
who measured a strong $a$-$b$ anisotropy in the electronic dispersion
of monocrystalline $\mbox{YBa}^{\ }_2\mbox{Cu}^{\ }_3\mbox{O}^{\ }_{6.993}$. 
In particular, they reported a 50 $\%$ difference in the magnitude
of the superconducting gap in the vicinity of the $(\pi,0)$ and
$(0,\pi)$ region of the first Brillouin zone (BZ), respectively. 
Smilde \textit{et al.}~\cite{smilde05} measured an $a$-$b$
anisotropy of the Josephson current in junctions between
monocrystalline $\mbox{YBa}^{\ }_2\mbox{Cu}^{\ }_3\mbox{O}^{\ }_7$ and $s$-wave
Nb, claiming that the obtained anisotropy can be well fitted by
a $83\%$ $d$-wave and $17\%$ $s$-wave order parameter.
Anisotropic responses are not limited to electromagnetic probes.
The dynamical magnetic susceptibility measured by INS 
in monocrystalline and fully detwinned YBCO
shows that the incommensurate peaks are strongly
anisotropic in that their line shapes and intensities break the
tetragonal symmetry.~\cite{stock04,stock05,mook,hinkov04}
Thus, it has become necessary to go beyond a pure $d^{\ }_{x^2-y^2}$
superconducting order parameter so as to incorporate the effects
of crystalline hosts with orthorhombic symmetry.

Strongly anisotropic INS responses have both been interpreted as
evidences for the proximity in parameter space to one-dimensional 
physics (stripe scenario) in Ref.~\onlinecite{mook}
or to two-dimensional physics (Fermi-liquid-like scenario)
in Ref.~\onlinecite{hinkov04}. 
The effects on INS of an orthorhombic dispersion of the
superconducting quasiparticles 
were previoulsy studied in Refs.~\onlinecite{Li04,Li05,manskeprl05,bascones05}.
In this article, we analyze the observed anisotropy in INS
within a conventional fermiology picture under the hypothesis 
that the observed anisotropies in the spin and charge response 
are caused by \textit{both} a subdominant $s$-wave component 
in the superconducting gap and an orthorhombic BCS dispersion. 
To this end we use a phenomenological
single-band tight-binding model
describing BCS quasiparticles interacting weakly through a residual
repulsive Hubbard interaction. 
The parameters entering the BCS dispersion 
are chosen so as to reproduce the measured values of the Fermi-surface 
and the BCS gaps at $(\pi,0)$ and $(0,\pi)$ close to optimally doped
YBCO.
The residual Hubbard interaction is fixed by the energy of the resonance
at $(\pi,\pi)$ at the same doping.~\cite{dai01}

The paper is organized as follows. Our model is described
in Sec.~\ref{sec: Definition of the fermiology model}.
Results for the dynamical magnetic susceptibility 
are presented in Sec.~\ref{sec: Numerical Results}.
The qualitative behavior of the 
dynamical magnetic susceptibility
is explained in Sec.~\ref{sec: discussions}.
We summarize with Sec.~\ref{sec: summary}.

\section{Definition of fermiology}
\label{sec: Definition of the fermiology model}

\begin{figure}[t]
\includegraphics[width=0.3\textwidth, angle=-0]{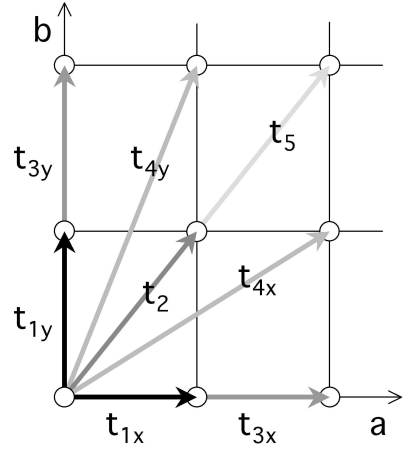}
\caption{
The hopping parameters used in the
tight-binding dispersion~%
(\protect\ref{eq: tight-binding dispersion})
are
$t^{\ }_1=-588.1$~meV,
$t^{\ }_2=146.1$~meV, 
$t^{\ }_3=9.5$~meV, 
$t^{\ }_4=-129.8$~meV, and
$t^{\ }_5=6.9$~meV
throughout this paper.
An orthorhombic symmetry implies that rotation symmetry
by $\pi/2$ is broken, i.e., that
$t^{\ }_{1x}\equiv t^{\ }_{1}(1+\delta^{\ }_{0})/4$
is not equal to
$t^{\ }_{1y}\equiv t^{\ }_{1}(1-\delta^{\ }_{0})/4$
and that
$t^{\ }_{3x}\equiv t^{\ }_{3}(1+\delta^{\ }_{0})/4$
is not equal to
$t^{\ }_{3y}\equiv t^{\ }_{3}(1-\delta^{\ }_{0})/4$.
         }
\label{Fig: hopping vectors}
\end{figure}

In this paper, we shall assume an effective 
one-band Hubbard Hamiltonian for each CuO$^{\ }_2$ plane
\begin{subequations}
\label{hubbard}
\begin{eqnarray}
H &=& H^{\ }_{0}+H^{\ }_{1},
\\
H^{\ }_{0} &=&
- \sum_{\langle ij \rangle' \sigma} 
t^{\ }_{ij} c^{\dag}_{i\sigma}c^{\ }_{j\sigma}
- \mu\sum_{i\sigma} n^{\ }_{i\sigma}
\nonumber\\
&&
- \sum_{\langle ij \rangle}
\left( 
\Delta^{\ }_{ij} c^{\dag}_{i \uparrow}c^{\dag}_{j \downarrow}
+ \textrm{h. c.} 
\right),
\\
H^{\ }_{1}&=&
U \sum_{i} n^{\ }_{i\uparrow} n^{\ }_{i\downarrow},
\label{eq: def residual interaction}
\end{eqnarray}
where the brackets $\langle i j \rangle$ and $\langle i j \rangle'$  denote the summation over the 
first nearest neighbors, and the first to fifth nearest-neighbors, respectively
(see Fig.~\ref{Fig: hopping vectors}).
Here,
$c^{\dag}_{i\sigma}$ 
is the creation operator of a quasiparticle with spin $\sigma$ on site $i$, 
$n^{\ }_{i\sigma} = c^{\dag}_{i\sigma} c^{\ }_{i\sigma}$
is the spin-dependent local number operator,
$t^{\ }_{ij}$ is a hopping matrix element in the CuO$^{\ }_2$ plane,
$\mu$ is the chemical potential,
$\Delta^{\ }_{ij}$ is the superconducting gap,
and $U$ denotes a residual on-site (i.e., intraorbital) Coulomb repulsion.
For simplicity we shall use a rigid-band approximation,
by which all the effects of doping
can be incorporated into a doping dependent chemical potential.
The summation over the first few nearest-neighbors pairs of directed sites
is most easily performed in the first BZ 
of the reciprocal space for the square lattice,
in which case the noninteracting Hamiltonian
is diagonal in reciprocal space
\begin{eqnarray}
H^{\ }_{0}=
-
\sum_{\boldsymbol{k}\in\mathrm{BZ}}
\left[
\varepsilon^{\ }_{\boldsymbol{k}}
\sum_{\sigma}
c^{\dag}_{\boldsymbol{k}\sigma} 
c^{\ }_{\boldsymbol{k}\sigma}
+
\Delta^{\ }_{\boldsymbol{k}}
\left(
c^{\dag}_{\boldsymbol{k}\uparrow} 
c^{\dag}_{\boldsymbol{k}\downarrow}
+
\mathrm{h. c.}
\right)
\right].
\nonumber\\
&&
\end{eqnarray}
We shall choose the band parameters so as to fit 
qualitatively the FS as measured by ARPES.
This can be done with the choice 
\begin{eqnarray}
\varepsilon^{\ }_{\bm{k}} & = & 
\frac{t^{\ }_1}{2} ( 1+ \delta^{\ }_0) \cos k^{\ }_x +
\frac{t^{\ }_1}{2}(1 - \delta^{\ }_0) \cos k^{\ }_y 
\nonumber\\
&&
+ t^{\ }_2 \cos k^{\ }_x \cos k^{\ }_y 
\nonumber\\
&&
+
\frac{t^{\ }_3}{2}( 1 + \delta^{\ }_0 )
\cos 2 k^{\ }_x + \frac{t^{\ }_3}{2}(1 - \delta^{\ }_0) \cos 2 k^{\ }_y 
\nonumber\\
&&
+ \frac{t^{\ }_4}{2} \cos 2 k^{\ }_x \cos k^{\ }_y 
+ \frac{t^{\ }_4}{2} \cos  k^{\ }_x \cos 2 k^{\ }_y 
\nonumber\\
&&
+ t^{\ }_5 \cos 2 k^{\ }_x \cos 2 k^{\ }_y + \mu.
\label{eq: tight-binding dispersion}
\end{eqnarray}
The values for the hopping matrix elements are those that
Norman used in Ref.~\onlinecite{norman01} to fit photoemission experiments.
The parameter $\delta^{\ }_{0}\neq0$ breaks the tetragonal symmetry
as the $k^{\ }_x$ and $k^{\ }_y$ directions in the BZ of the square lattice are
not equivalent. In this paper we shall always choose a nonvanishing
$\delta^{\ }_{0}<0$ that corresponds to \textit{effective hopping amplitudes
larger along the $k^{\ }_{y}$ direction than along the $k^{\ }_{x}$ direction}.
The superconducting gap is also chosen, on 
phenomenological grounds and out of simplicity, to be
\begin{eqnarray}
\Delta^{\ }_{\bf{k}} = 
\left( \Delta^{\ }_{x} \cos k^{\ }_x - \Delta^{\ }_{y}\cos k^{\ }_y \right) /2 
+ \Delta^{\ }_s, 
\label{eq:swavegap}
\end{eqnarray}
where 
\begin{eqnarray}
0<\Delta^{\ }_s< 
\Delta^{\ }_0\equiv\left(\Delta^{\ }_{x}+\Delta^{\ }_{y}\right)/2.
\end{eqnarray}
\end{subequations}
The condition 
$| \Delta^{\ }_{(\pi,0)} | < | \Delta^{\ }_{(0, \pi)}|$ 
that is observed in ARPES (see Ref.\ \onlinecite{lu01})
can be implemented with the choice 
$\Delta^{\ }_{0}\equiv\Delta^{\ }_{x}=\Delta^{\ }_{y}$ and $\Delta^{\ }_s > 0$
for the effective gap parameters.
Of course, this choice is not unique, but since we are not concerned with
deducing in a self-consistent manner the band and gap parameters from a microscopic
model, we will make it for simplicity.
In general, the effective energy scale $U$ can encode an interaction 
that is strongly momentum dependent.
For example in the $1/z$ expansion with $z$ as the number of nearest neighbors,
the repulsive channel of the interaction is peaked at
the AF wave vector ${\bf Q}^{\ }_{AF}=(\pi,\pi)$.~\cite{si,brinckmann01,yamase01,onufrieva}
This, however, will have no bearing on our conclusions and we choose $U$
to represent a Hubbard on-site repulsion out of simplicity.
The value for $U$ throughout this paper 
is fixed by demanding that the position in energy of
the resonance at the wave vector $(\pi,\pi)$
coincides with the one observed
in optimally doped YBCO.

\begin{figure}[t!]
\vspace{2 mm}
\includegraphics[width=0.47\textwidth]{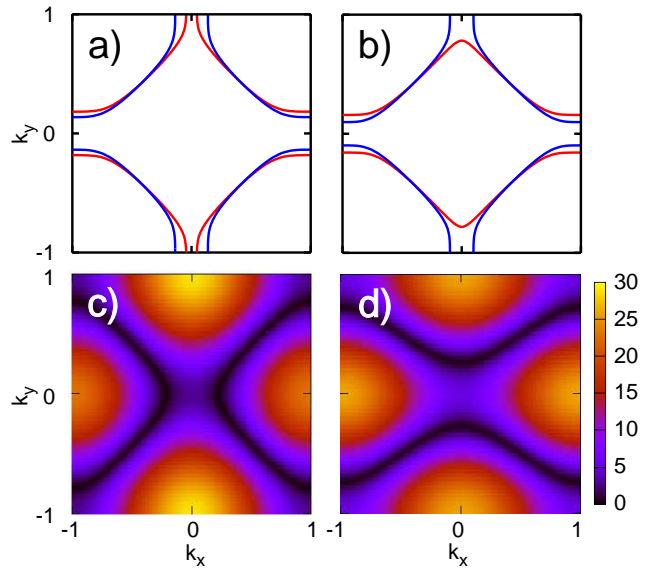}
\caption{(Color online) 
Panels (a) and (b) display the FS for the
tight-binding dispersion~\re{eq: tight-binding dispersion}
with an orthorhombic distortion 
$\delta^{\ }_0 = 0$ (blue) 
and
$\delta^{\ }_0=-0.03$ (red) 
using two different values of
the chemical potential 
$\mu=110$ and $\mu=120$ meV, respectively.
The numerical values taken by the 5 hopping parameters
$t^{\ }_{1},\ldots,t^{\ }_{5}$ 
are given in Fig.~\ref{Fig: hopping vectors}.
Panel (c) displays the absolute value of the superconducting gap
$\Delta^{\ }_{\bm{k}}=  
\Delta^{\ }_0 \left( \cos k^{\ }_x - \cos k^{\ }_y\right)/2 
+ \Delta^{\ }_s$ in meV using $\Delta^{\ }_0 = 26$ meV
and $\Delta^{\ }_s=3$ meV. Panel (d) displays the absolute value of the
superconducting gap
$\Delta^{\ }_{\bm{k}}= 
\left(\Delta^{\ }_x \cos k^{\ }_x -  \Delta^{\ }_y \cos k^{\ }_y\right)/2$ 
with $\Delta^{\ }_x = 20.8$ meV and $\Delta^{\ }_y=31.2$ meV.} 
\label{Fig: fermi surface}
\end{figure}

The shape of the FS is depicted in 
Fig.~\ref{Fig: fermi surface}(a) 
for the band parameters (see Ref.~\onlinecite{norman01}), 
$\mu=110.0$~meV, 
$t^{\ }_1=-588.1$~meV,
$t^{\ }_2=146.1$~meV, 
$t^{\ }_3=9.5$~meV, 
$t^{\ }_4=-129.8$~meV, and
$t^{\ }_5=6.9$~meV for 
$\delta^{\ }_0 = 0$ (blue) 
and $\delta^{\ }_0=-0.03$ (red). 
The Fermi arcs of the orthorhombic FS are
closer together in the $(0,\pm\pi)$ region than in
the $(\pm\pi,0)$ region of the BZ. 
This is a consequence of taking $\delta^{\ }_0 < 0$.
The opposite result follows from the choice $\delta^{\ }_0 > 0$.%
\cite{footnote}
It is possible to use the chemical potential $\mu$ as a
tuning parameter through a phase transition of the 
Fermi surface topology. The FS in
Fig.~\ref{Fig: fermi surface}(a) 
is two-dimensional and holelike, 
i.e., it is closed around the four corners of the
first BZ of the square lattice. 
Upon increasing the chemical potential to the value 
$\mu= 120$~meV, 
the FS~[Fig.~\ref{Fig: fermi surface}(a)]  
undergoes a transition to the quasi-one-dimensional
topology~[Fig.~\ref{Fig: fermi surface}(b)]
by which it is now open along the $k^{\ }_{x}$ direction 
but closed along the $k^{\ }_{y}$ direction in the first BZ 
of the square lattice. 
It has been argued in Refs.~\onlinecite{yamase01} and~\onlinecite{metzner}
that such a distorted FS can arise as a result of a
$d^{\ }_{x^2-y^2}$-wave Pomeranchuk instability due to strong
electron-electron interactions.
The absolute value of the superconducting gap
$
\Delta^{\ }_{\bm{k}}=  
\Delta^{\ }_0 \left( \cos k^{\ }_x - \cos k^{\ }_y\right)/2 + \Delta^{\ }_s
$  
as a function of $\bm{k}$ with
$\Delta^{\ }_0 = 26$ meV and $\Delta^{\ }_s=3$ meV
is shown in Fig.~\ref{Fig: fermi surface}(c).
The nodal points form two lines 
that are closed around the points
$(\pm\pi,0)$, respectively, in the extended BZ.
The choice
$
\Delta^{\ }_{\bm{k}}=  
\left( 
\Delta^{\ }_{x}\cos k^{\ }_x - \Delta^{\ }_{y} \cos k^{\ }_y\right)/2 + \Delta^{\ }_s
$  
as a function of $\boldsymbol{k}$
with 
$\Delta^{\ }_{x}=20.8$ meV,
$\Delta^{\ }_{y}=31.2$ meV,
and
$\Delta^{\ }_{s}=0$
(extended $s$-wave subdominant component)
is shown in~Fig.~\ref{Fig: fermi surface}(d).
A subdominant extended $s$-wave component with 
$\Delta^{\ }_{y} > \Delta^{\ }_{x}$ was found
in Refs.~\onlinecite{Li05, yamase06}
after solving self-consistently a $t-t^{\prime}-J$ model
treated by the slave-boson approach.

In this paper we shall approximate the full
frequency $\omega$
and momentum $\boldsymbol{q}$-dependent dynamical spin susceptibility
$\chi(\omega,\boldsymbol{q})$
by the RPA approximation in terms of the
noninteracting BCS-Lindhard response function
$\chi^{\ }_{0}(\omega,\boldsymbol{q})$.
In turn, as the INS intensity 
in the superconducting state is proportional to
the imaginary part 
$\chi^{\prime\prime}(\omega,\boldsymbol{q})$
of 
$\chi(\omega,\boldsymbol{q})$, 
we shall be computing 
\begin{equation}
\chi^{\prime\prime}_{\hbox{\tiny RPA}} (\omega,\boldsymbol{q}) = 
\frac{
\chi^{\prime\prime}_{0}(\omega,\boldsymbol{q})
     }
     {
\left[
1-U\chi^{\prime}_{0}(\omega,\boldsymbol{q})
\right]^2 
+ 
U^2\chi^{\prime\prime2}_{0}(\omega,\boldsymbol{q})
     }.
\label{eq: chi RPA}
\end{equation}
A dispersing branch of incommensurate or commensurate peaks 
occurs whenever it is possible to find a frequency-momentum pair
$(\omega^*,\boldsymbol{q}^*)$
that satisfies the dynamical Stoner criterion
\begin{equation} 
1-U\chi^{\prime}_{0}(\omega^{*},\boldsymbol{q}^{*}) = 0.
\label{ucr}
\end{equation}
The height of the peaks
in a momentum or energy scan is determined by
the size of 
$\chi^{\prime\prime}_{0}(\omega^{*},\boldsymbol{q}^{*})$ 
is.
We define the resonance energy $\omega^{\ }_{res}$ as $\omega^{\ast}$ 
at the AF wave vector $\bm{q}^{\ast}=(\pi, \pi)$.
It is of order $43$ meV for the band parameters of 
Fig.~\ref{Fig: hopping vectors},
the arithmetic average gap maximum $\Delta^{\ }_{0}$
taking the value of $26$ meV, and the choice $U=155$ meV.

%
\section{Numerical Results for the dynamical magnetic susceptibility}
\label{sec: Numerical Results} 

We have computed numerically the imaginary part of the RPA
spin susceptibility~\re{eq: chi RPA} at a
fixed transfer energy as a function of $\bf{q}$ 
for values of the transfer energy ranging from well below
to well above the resonance energy $\sim 43$ meV.
The band parameters in Fig.~\ref{Fig: hopping vectors}
and the arithmetic average gap maximum $\Delta^{\ }_{0}=26$ meV
are fixed throughout this section.
The values taken by 
the subdominant $s$-wave component $\Delta^{\ }_{s}$
and the orthorhombic parameters $\delta^{\ }_{0}$ 
and $|\Delta^{\ }_{x}-\Delta^{\ }_{y}|$
are varied.

\begin{figure}[t]
\vspace{2 mm}
\includegraphics[width=0.47\textwidth, angle=-0]{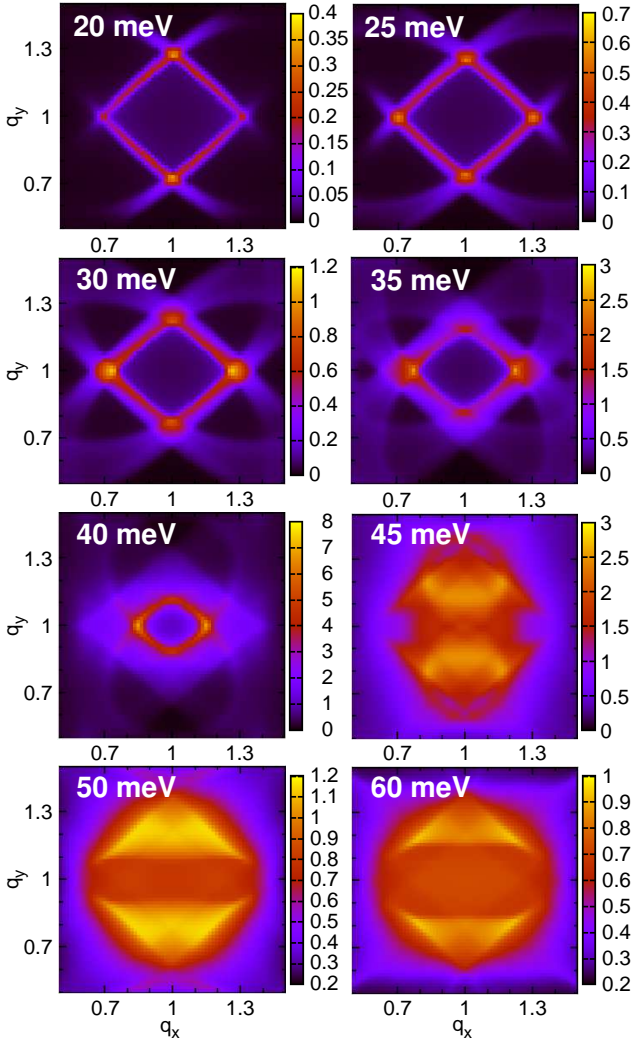}
\caption{(Color online) 
Imaginary part of the RPA spin susceptibility
$\chi^{\prime\prime}_{\hbox{\tiny RPA}}(\omega,\boldsymbol{q})$  
for a constant transfer energy 
$\omega = 20\mbox{meV},\ldots,60$meV as a function
of $\boldsymbol{q}$ (in units of $\pi$) for the tight-binding band
structure of Fig.~\ref{Fig: fermi surface}(a) with 
$\delta^{\ }_0=-0.03$. 
We are also using 
$\Delta^{\ }_0=26$ meV, 
$\Delta^{\ }_s=3$ meV, 
$U=155$ meV, 
$T=0$ K,
and a damping 
$\Gamma=1$ meV.
        }
\label{Fig: im chi q map with full orthorhombic}
\end{figure}

(i) \textit{FS with orthorhombic anisotropy. Gap with isotropic 
$s$-wave subdominant component.}
The case of a weakly orthorhombic distorted FS and of
an orthorhombic gap induced by a weak $s$-wave subdominant component
is displayed in Fig.~\ref{Fig: im chi q map with full orthorhombic}.
The band structure corresponds to that in 
Fig.~\ref{Fig: fermi surface}(a)
with $\delta^{\ }_0 = -0.03$ 
and the anisotropic gap 
of Fig.~\ref{Fig: fermi surface}(c), i.e.,
$|\Delta^{\ }_{x}-\Delta^{\ }_{y}|=0$
while $\Delta^{\ }_s=3$ meV.
Most of the intensity in 
$\chi^{\prime\prime}_{\hbox{\tiny RPA}}(\omega,\boldsymbol{q})$
is concentrated on the perimeter of a diamond
that is centered around the AF wave vector $(\pi,\pi)$ for energies
smaller than $40$ meV. The area enclosed by this diamond decreases 
with increasing transfer energies. 
Remarkably, the maximum intensity
is on the upper and lower corners of the diamond 
[intersection between the diamond and the vertical line passing through
$(\pi,\pi)$]
at the transfer energy of $20$ meV whereas it has moved to the
left and right corners of the diamond
[intersection between the diamond and the horizontal line passing through
$(\pi,\pi)$]
at the transfer energy of $30$ meV.
The ratio of intensities at the upper and left corners of the diamond
is of order 2 (1/2) for the transfer energy of
$20$ meV ($35$ meV).
This anisotropy is much stronger than the 
orthorhombic anisotropy in the dispersion of the BCS quasiparticles
(a 10$\%$ effect induces a 100$\%$ effect).
For comparison,
one finds that most of the intensity in 
$\chi^{\prime\prime}_{\hbox{\tiny RPA}}(\omega,\boldsymbol{q})$
is to be found in a ring centered around $(\pi,\pi)$ with four
pronounced peaks at 
$(\pi\pm q^{\ }_0, \pi)$ 
and
$(\pi, \pi\pm q^{\ }_0)$
in the tetragonal case, $\delta^{\ }_0=\Delta^{\ }_s=0$ 
[not shown here, see Fig. 4(a) in Ref.~\onlinecite{manskeprl05}].
For energies
larger than $40$ meV the 
intensity in 
$\chi^{\prime\prime}_{\hbox{\tiny RPA}}(\omega,\boldsymbol{q})$
is suppressed along the $x$ axis passing through $(\pi, \pi)$
and is mostly
 concentrated in a disc 
that is centered around the AF wave vector.

\begin{figure}[t!]
\vspace{2 mm}
\includegraphics[width=0.23\textwidth, angle=-0]{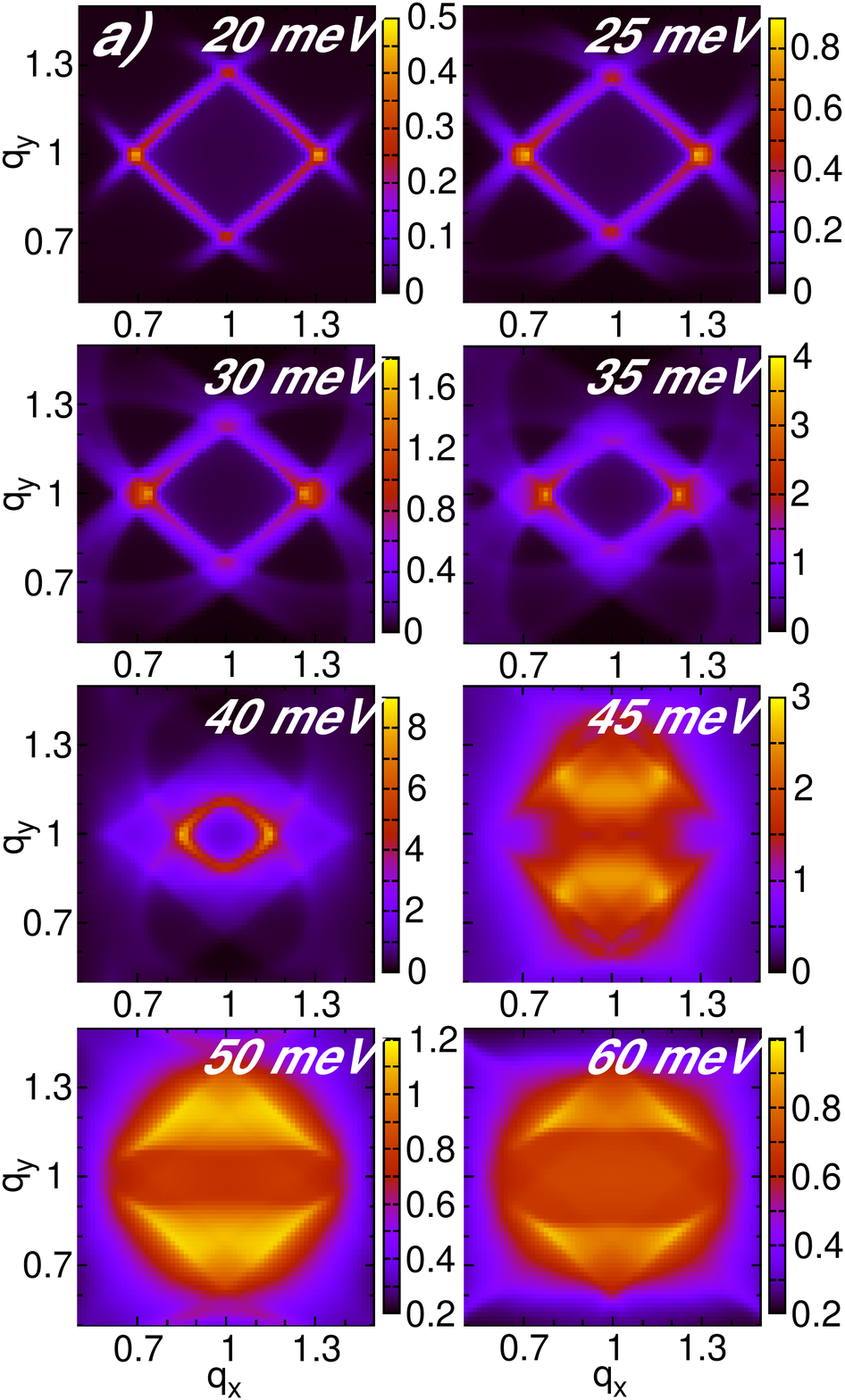}
\includegraphics[width=0.23\textwidth, angle=-0]{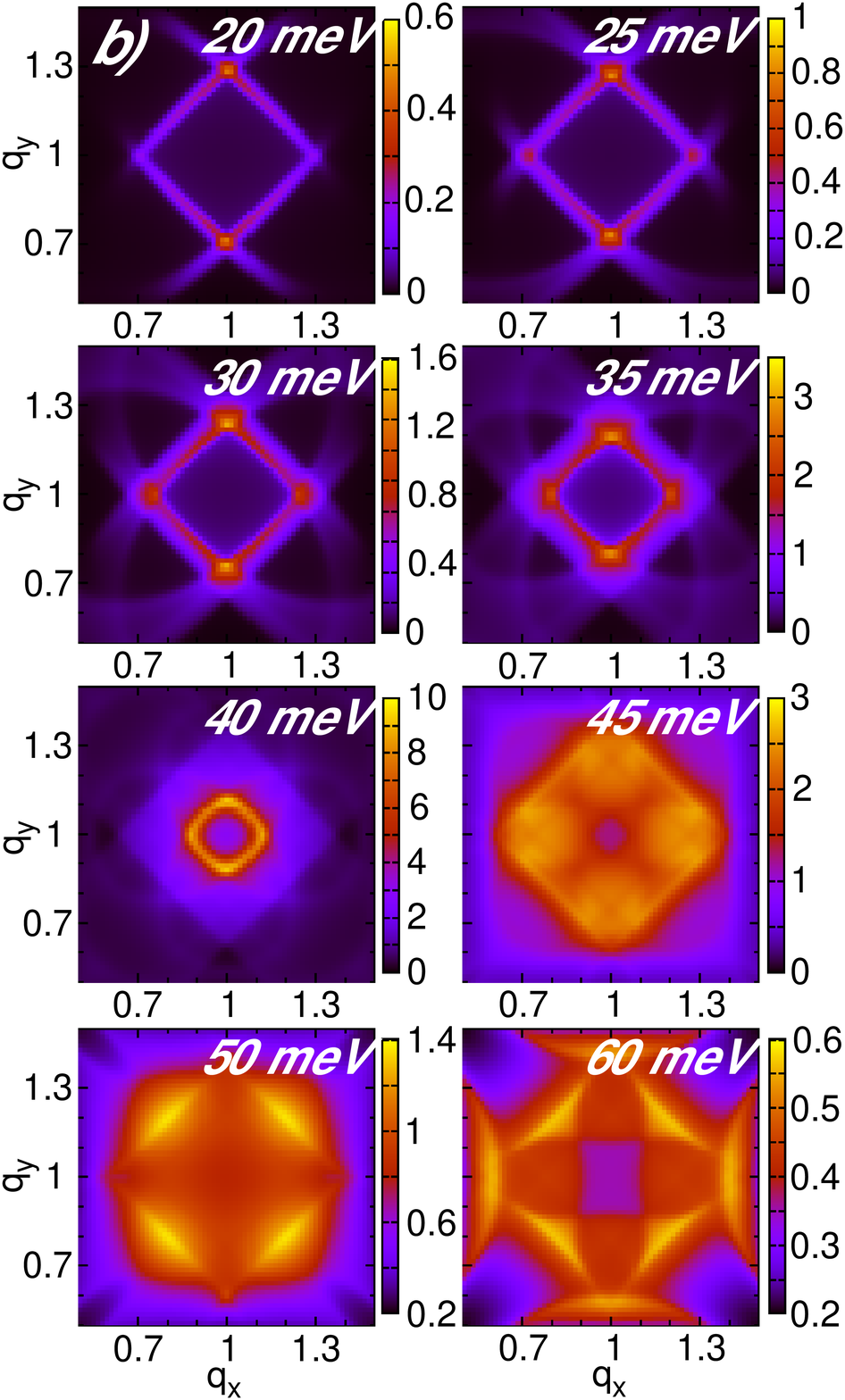}
\caption{(Color online)
(a) Same as in Fig.~\ref{Fig: im chi q map with full orthorhombic}
except for $\Delta^{\ }_s=0$. 
(b) Same as in Fig.~\ref{Fig: im chi q map with full orthorhombic}
except for $\delta^{\ }_0=0$.
        }
\label{Fig: im chi q if gap is tetragonal}
\label{Fig: figure4}
\end{figure}

(ii) \textit{FS with orthorhombic anisotropy. Gap with tetragonal symmetry.}
The case of a weakly orthorhombic distorted FS,
$\delta^{\ }_0 = -0.03$, and of
a tetragonal gap, $|\Delta^{\ }_{x}-\Delta^{\ }_{y}|=\Delta^{\ }_s=0$,
 is displayed in Fig.~\ref{Fig: figure4}(a).
The only qualitative difference with 
Fig.~\ref{Fig: im chi q map with full orthorhombic}
is the fact that the maximum intensity is always found
at the left and right corners of a diamond centered at $(\pi,\pi)$
below the resonance energy $\sim43$ meV. Evidently,
this difference at the lower end of the transfer energies $\sim 20$ meV
can be ascribed to switching off the $s$-wave subdominant component
to the gap.
We thus conclude that, below the resonance energy,
the anisotropy $t^{\ }_{y}/t^{\ }_{x}>1$ 
favors dominant incommensurate peaks along the $q^{\ }_{x}$ direction 
while the subdominant $s$-wave component
$\Delta^{\ }_{s}>0$ favors 
dominant incommensurate peaks along the $q^{\ }_{y}$ direction.

(iii) \textit{FS with tetragonal symmetry. Gap with isotropic $s$-wave subdominant component.}
The case of a tetragonal FS, $\delta^{\ }_0 = 0$, and of
an orthorhombic gap induced by a weak $s$-wave subdominant component,
$\Delta^{\ }_s=3$ meV,
is displayed in Fig.~\ref{Fig: figure4}(b).
There are two qualitative differences with
Fig.~\ref{Fig: im chi q map with full orthorhombic}.
The maximum intensity is always found
at the upper and lower corners of a diamond centered at $(\pi,\pi)$
below the resonance energy  $\sim43$ meV.
The intensity distribution above $\sim43$ meV is much less
blurry than in Fig.~\ref{Fig: figure4}(a)
and displays some well defined arcs of dominant intensity
centered about the diagonals passing through the center $(\pi,\pi)$
of the magnetic BZ.

\begin{figure}[t]
\vspace{2 mm}
\includegraphics[width=0.23\textwidth, angle=-0]{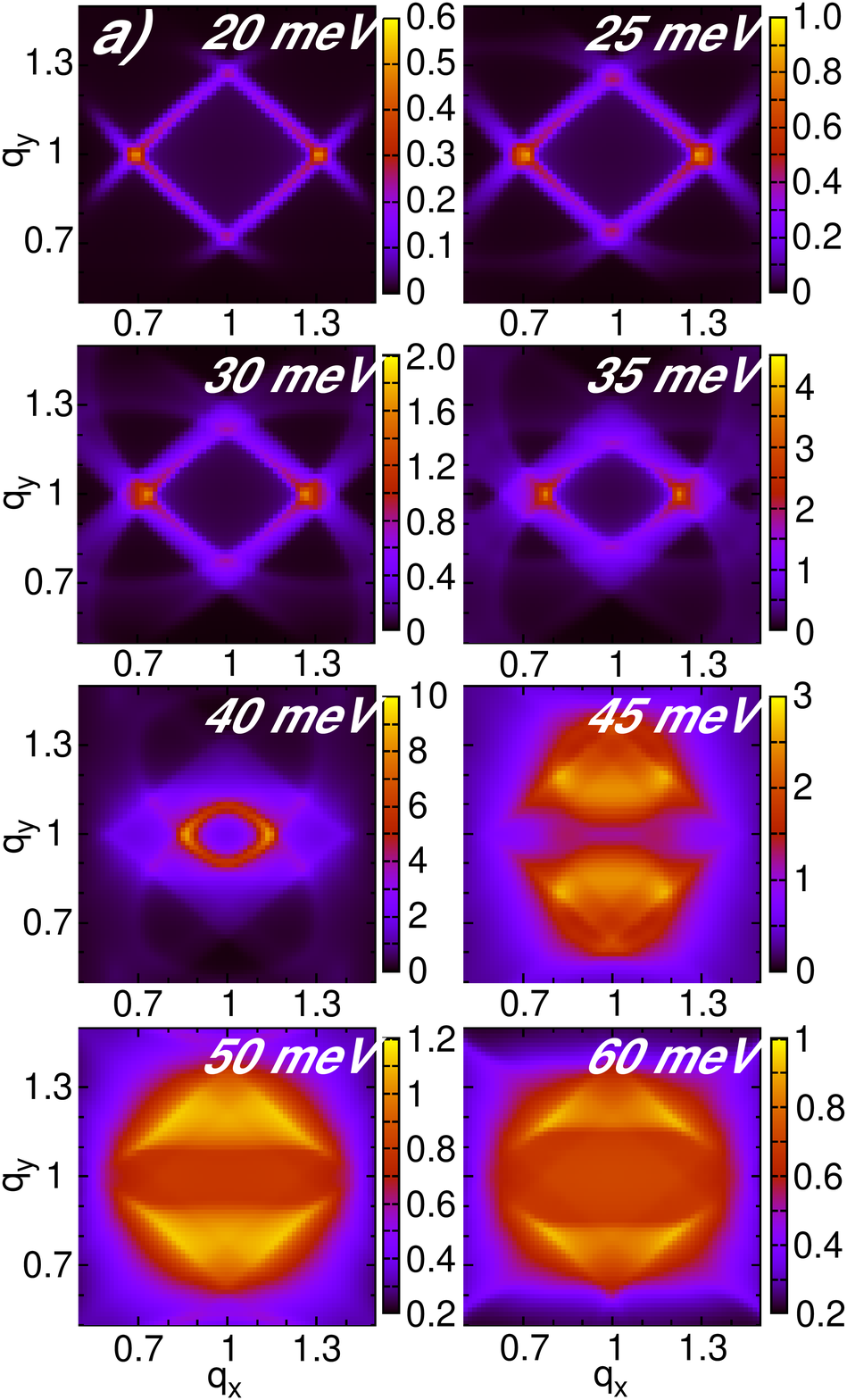}
\includegraphics[width=0.23\textwidth, angle=-0]{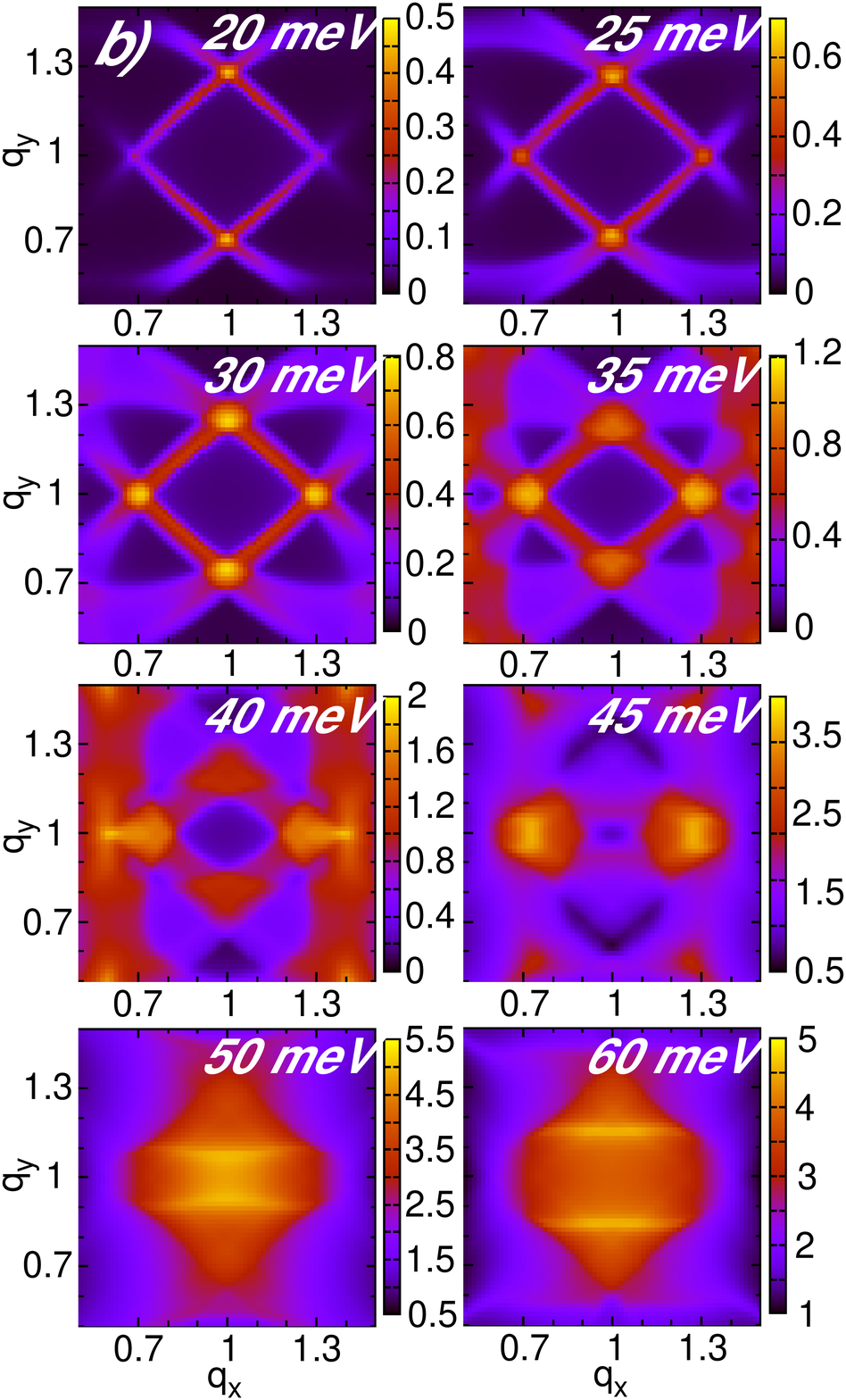}
\hfill
\caption{(Color online)
(a) Same as in Fig.~\ref{Fig: im chi q map with full orthorhombic}
except for the gap
$\Delta^{\ }_{\bm{k}}= 
\left(\Delta^{\ }_x \cos k^{\ }_x -  \Delta^{\ }_y \cos k^{\ }_y\right)/2$ 
with $\Delta^{\ }_x = 20.8$ meV and $\Delta^{\ }_y=31.2$ meV
from Fig.~\ref{Fig: fermi surface}(d).
(b)
Imaginary part of the BCS-Lindhard spin susceptibility
$\chi^{\prime\prime}_{0}(\omega,\boldsymbol{q})$  
for a constant transfer energy 
$\omega = 20\mbox{meV},\ldots,60$meV as a function
of $\boldsymbol{q}$ (in units of $\pi$) for 
the same parameters as in Fig.~\ref{Fig: im chi q map with full orthorhombic}.
        }
%
\label{Fig: figure5}
\end{figure}

(iv) \textit{FS with orthorhombic anisotropy. Gap with extended $s$-wave
subdominant component.}
The case of a weakly orthorhombic distorted FS, $\delta^{\ }_0 = -0.03$, and of
a gap with a strong orthorhombic distortion induced by
$|\Delta^{\ }_{x}-\Delta^{\ }_{y}|=10.4$ meV ($\Delta_y > \Delta_x$),
 but $\Delta^{\ }_s=0$,
is displayed in Fig.~\ref{Fig: figure5}(a).
It is qualitatively very similar to
Fig.~\ref{Fig: im chi q if gap is tetragonal}(a).
The maximum intensity is always found
at the left and right corners of a diamond centered at $(\pi,\pi)$
below the resonance energy $\sim43$ meV.
The anisotropy in the ratio between the intensities at
the upper and left corners of the diamond are more pronounced than
in Fig.~\ref{Fig: im chi q if gap is tetragonal}(a).
Since the anisotropy in the hopping parameters dominates
over the anisotropy in the SC gap function, 
the opposite choice $\Delta_x > \Delta_y$ (not shown) leads to a qualitatively 
similar result, albeit with a reduced anisotropy ratio at low
transfer energies $\sim 20$~meV.

At last we illustrate with Fig.~\ref{Fig: figure5}(b)
the fact that the distribution of intensities below the resonance energy 
$\sim43$ meV
in the RPA spin susceptibility tracks that
in the bare Lindhard spin susceptibility.
It is in this sense that the qualitative evolution of the intensity distribution in
Fig.~\ref{Fig: im chi q map with full orthorhombic}
between 20 and 35 meV is robust to
changing the momentum dependence of the residual 
quasiparticle interaction in Eq.~(\ref{eq: def residual interaction}).

\section{Discussions}
\label{sec: discussions} 

\begin{figure}[t!]
\includegraphics[width=0.47\textwidth, angle=0]{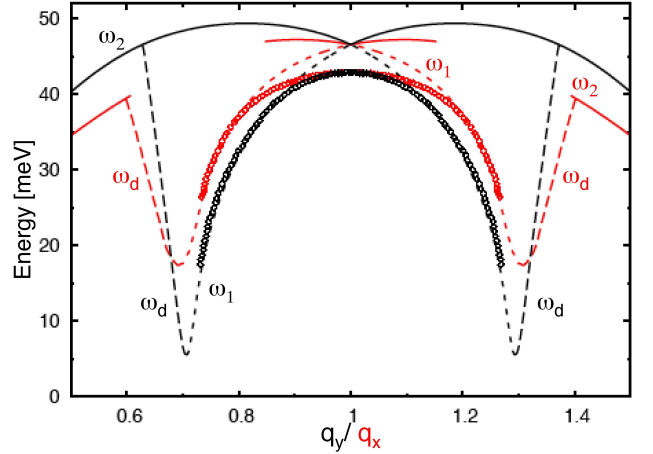}
\caption{(Color online)
Momentum dependence of the threshold frequencies 
$\omega^{\ }_1$, $\omega^{\ }_2$, and $\omega^{\ }_d$ calculated from
$\min_{\bf{k}} E^{\ }_2(\bf{q},\bf{k})$ 
using the same parameters as
in Fig. \protect\ref{Fig: im chi q map with full orthorhombic}. 
The threshold frequency as a function of
$(\pi,q^{\ }_y)$ ($(q^{\ }_x,\pi)$) is depicted in black (red).
The open diamonds represent the position of the resonance peak.
        }
\label{Fig: fulldisp}
\end{figure}

In this section we explain the qualitative behavior of
the imaginary part of the RPA spin susceptibility
$\chi^{\prime\prime}_{\hbox{\tiny RPA}}(\omega, \bm{q})$ 
for an orthorhombic superconductor
 in terms of  the properties of $\chi^{\prime\prime}_{0}(\omega, \bm{q})$ and the
two-particle energy $E^{\ }_2(\bm{q}, \bm{k})$.
We recall that in the limit of $T=0$ and for positive frequencies
 the imaginary part of the noninteracting
BCS-Lindhard response function $\chi^{\ }_0 ( \omega, \bm{q})$
simplifies to~\cite{schnyder04}
\begin{eqnarray} \label{eq: chi''_0 at T=0}
\chi^{\prime\prime}_0 ( \omega, \bm{q} ) 
& = &
\frac{\pi}{N}
\sum_{\bm{k}} 
C^{+,-}_{\bm{q}, \bm{k}} 
\delta \big(  \omega- E^{\ }_2(\bm{q}, \bm{k}) \big),
\\
C^{+,-}_{\bm{q}, \bm{k}}
&=&
\frac{1}{4} 
\left( 1 - \frac{\varepsilon^{\ }_{\bm{k}+\bm{q}}
\varepsilon^{\ }_{\bm{k}} + \Delta^{\ }_{\bm{k}+\bm{q}} \Delta^{\ }_{\bm{k}}}
{E^{\ }_{\bm{k}+\bm{q}} E^{\ }_{\bm{k}} } 
\right) ,
\\
E^{\ }_2 ( \bm{q}, \bm{k} )
&=&
E^{\ }_{\bm{k}+ \bm{q}} + E^{\ }_{\bm{k}},
\end{eqnarray}
where 
$E^{\ }_{\bm{k}} = \sqrt{ \varepsilon_{\bm{k}}^2 + \Delta_{\bm{k}}^2}$ 
denotes the dispersion of the quasiparticles in the superconducting state.
At a fixed wave vector $\bm{q}$ the imaginary part of the noninteracting 
spin susceptibility $\chi^{\prime\prime}_0(\omega, \bm{q})$ vanishes below the threshold
frequency 
\begin{eqnarray}
\omega^{\ }_{c}(\bm{q})=
\min_{\bm{k}\in\mathrm{BZ}} E^{\ }_2(\bm{q},\bm{k})
\end{eqnarray}
that defines the border to a continuum of particle-hole excitations.
For a $d$-wave superconductor the low-energy border of the continuum 
has a nontrivial form (see Fig.~\ref{Fig: fulldisp}). 
It is bounded by several segments of different curves 
along each of which $\chi^{\prime\prime}_0 ( \omega, \bm{q})$ exhibits 
either a jump ($\omega^{\ }_1$ and $\omega^{\ }_2$ in Fig.~\ref{Fig: fulldisp})
or a kink ($\omega^{\ }_d$ in Fig.~\ref{Fig: fulldisp})
as a function of frequency, depending on whether the coherence factor
$C^{+,-}_{\bm{q}, \bm{k}}$ in Eq.~\re{eq: chi''_0 at T=0} is vanishing 
for the wave vectors $\bm{k}$ contributing to $\chi^{\prime\prime}_0(\omega, \bm{q})$ at
the border to the continuum.~\cite{onufrieva} The size of the jump in 
$\chi^{\prime\prime}_0(\omega, \bm{q})$
is controlled by two criteria: (i) How flat the two-particle dispersion at the
corresponding minimum in $E^{\ }_2(\bm{q}, \bm{k})$ is, 
and (ii) by the degeneracy of the minimum itself. As explained in 
Ref.~\onlinecite{schnyder04} the degeneracy of the minima
$\min_{\bm{k}} E^{\ }_2 ( \bm{q}, \bm{k})$
is increased  for $\bm{q}$ on a high symmetry axes of the magnetic BZ, 
i.e., on 
the $k^{\ }_x$- or $k^{\ }_y$-axes passing through $(\pi, \pi)$
in the case of orthorhombic symmetry.
%
The dispersion of the spin excitations in the presence of 
interactions is to a large extent determined by
the behavior of $\chi^{\prime\prime}_0 (\omega, \bm{q})$
at the border to the particle-hole continuum.
A steplike discontinuity in the frequency dependence
of $\chi^{\prime\prime}_0(\omega, \bm{q})$ results in a logarithmic singularity in
$\chi^{\prime}_0(\omega, \bm{q})$  due to the Kramers-Kronig relation.
This in turn leads to a pole in 
$\chi^{\prime\prime}_{\hbox{\tiny RPA}}(\omega, \bm{q})$
since the dynamical Stoner criterion~\re{ucr} can be satisfied 
at a frequency $\omega^{\ast}(\bm{q}) < \omega^{\ }_c(\bm{q})$.
A finite damping $\Gamma$ cuts off the logarithmic singularity in
$\chi^{\prime}_0(\omega, \bm{q})$, and the 
dynamical Stoner criterion can only be met for a sufficiently large
size of the step in $\chi^{\prime\prime}_0(\omega, \bm{q})$ (see open diamonds
in Fig.~\ref{Fig: fulldisp}).

We find that an orthorhombic distortion  in the band structure  or in
the superconducting order parameter partially lifts 
the degeneracy of the minima in $E^{\ }_2(\bm{q}, \bm{k})$ for $\bm{q}$ on the 
diagonal axes passing through $(\pi, \pi)$. That is, for orthorhombic
symmetry and $\bm{q}$ on the diagonal lines, there are four twofold 
degenerate critical frequencies $\omega^{\ }_i(\bm{q})$ along which 
$\chi^{\prime\prime}_0(\omega, \bm{q})$ exhibits a jump. 
Whereas in the tetragonal 
case there are one fourfold and two twofold 
degenerate threshold frequency $\omega^{\ }_i(\bm{q})$.
Consequently, the intensity maxima in 
$\chi^{\prime\prime}_{\hbox{\tiny RPA}}(\omega, \bm{q})$, 
for  $\omega < \omega^{\ }_{res}$ and for orthorhombic symmetry, lie on the 
horizontal and vertical axes passing through $(\pi, \pi)$. 
This is in contrast to the tetragonal case, where the intensity maxima can
occur on the diagonal axes as well.

\begin{figure}[t!]
\vspace{2 mm}
\includegraphics[width=0.47\textwidth, angle=0]{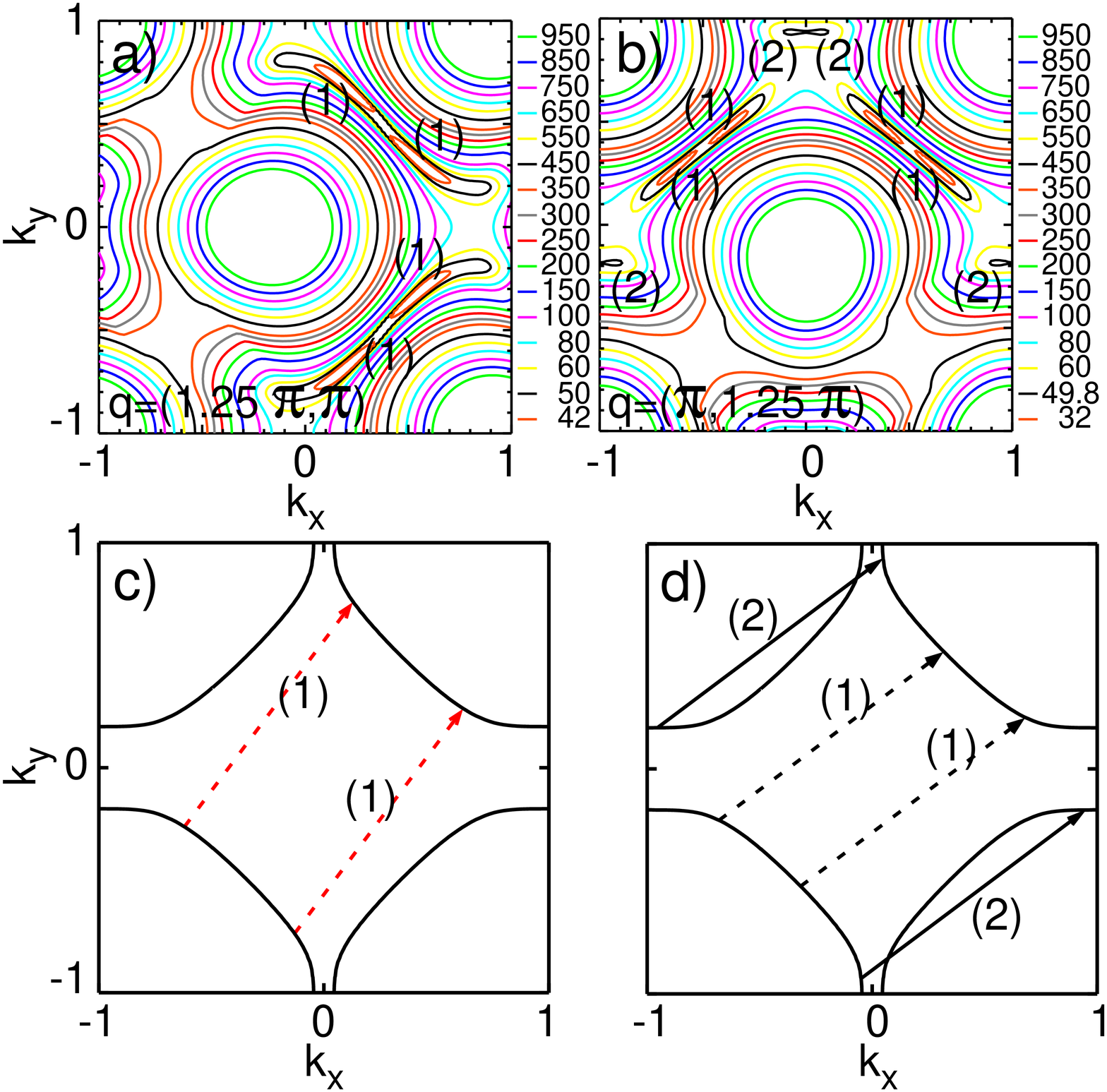}
\hfill
\caption{(Color online) 
Panel (a) and (b) display the calculated
$\bm{k}$-dependence of $E^{\ }_2(\bm{q},\bm{k})$ 
in the first BZ 
for $\bm{q}=(1.25\pi,\pi)$ and $\bm{q}=(\pi,1.25\pi)$, 
respectively,
using the same parameters as in Fig.~\
\protect\ref{Fig: im chi q map with full orthorhombic}. 
Panel (c) and (d) show the transitions between points on the Fermi surface
yielding the threshold frequencies 
$\omega^{\ }_1$ and $\omega^{\ }_2$
for $\bm{q}=(1.25\pi,\pi)$ and $\bm{q}=(\pi,1.25\pi)$, 
respectively.
        }
\label{Fig: contour}
\end{figure}

\begin{figure}[t!]
\includegraphics[width=0.47\textwidth, angle=0]{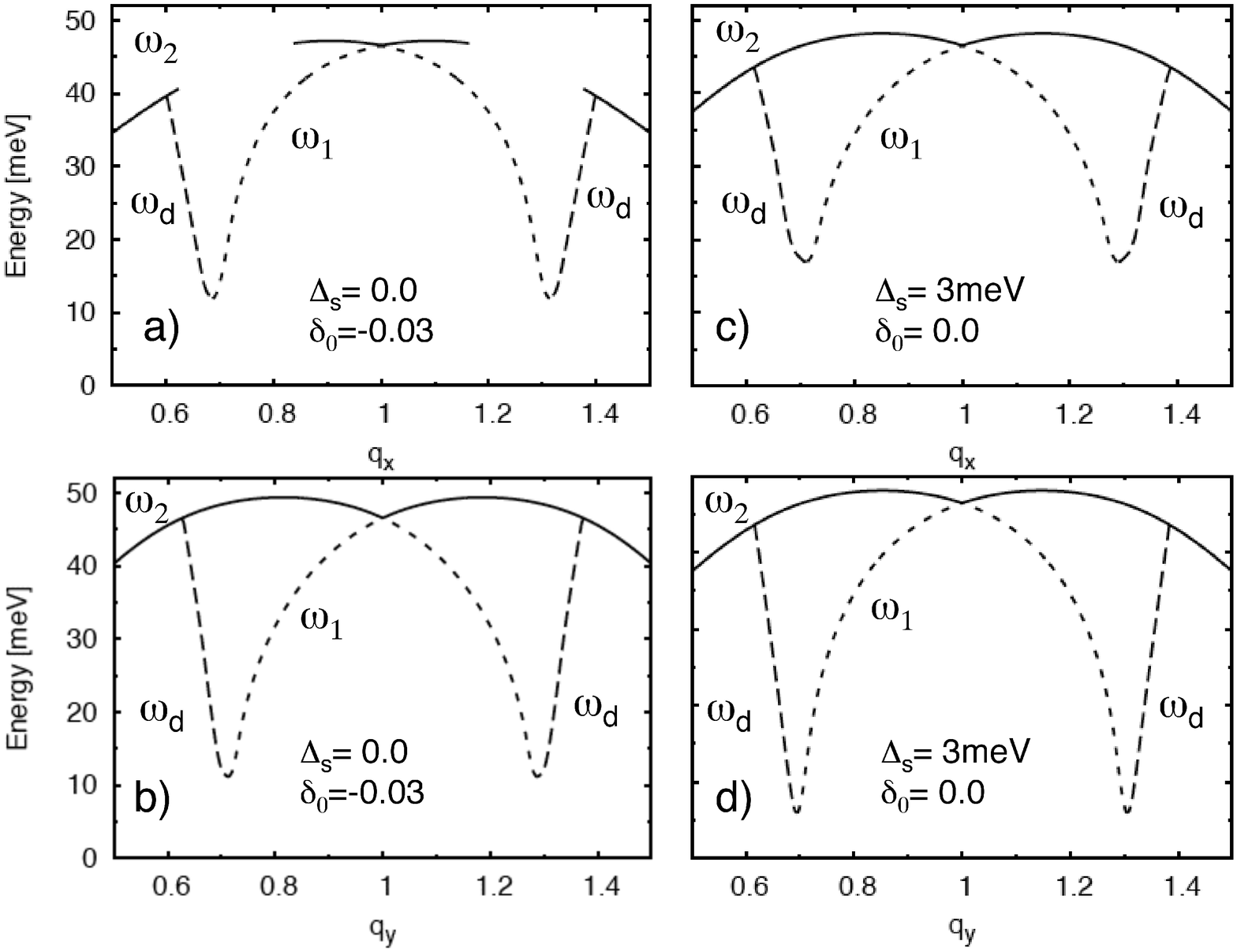}
\caption{
Momentum dependence of the threshold
frequencies $\omega^{\ }_1$, $\omega^{\ }_2$, and $\omega^{\ }_d$ 
calculated from
$\min_{\bf{k}} E^{\ }_2(\bf{q},\bf{k})$ 
for different FS and gap parameters
with $\bm{q} =(q^{\ }_x,\pi)$ in panels (a) and (c) 
while $\bm{q} =(\pi,q^{\ }_y)$ in panels (b) and (d). 
The FS parameters are those of 
Fig.~\ref{Fig: fermi surface}(a) with $\delta^{\ }_{0}=-0.03$
while the gap is a pure $d$-wave gap with
$\Delta^{\ }_0=26$~meV and $\Delta^{\ }_s=0$
in both panels (a) and (b).
The FS parameters are those of 
Fig.~\ref{Fig: fermi surface}(a) with $\delta^{\ }_{0}=0$
while the gap parameters are those of 
Fig.~\ref{Fig: fermi surface}(c) with $\Delta^{\ }_{s}=3$~meV
in both panels (c) and (d).
        }
\label{Fig: sepdisp}
\end{figure}

In Fig.~\ref{Fig: fulldisp} we present the electron-hole continuum
and the threshold frequencies $\omega^{\ }_1(\bm{q})$, $\omega^{\ }_2(\bm{q})$,
and $\omega^{\ }_d(\bm{q})$
 for the directions $(q^{\ }_x, \pi)$  and $(\pi, q^{\ }_y)$ 
using the same parameters as in 
Fig.~\ref{Fig: im chi q map with full orthorhombic}. 
Also shown is the continuation
of the threshold lines $\omega^{\ }_2(\bm{q})$ into the continuum
along which $\chi^{\prime\prime}_0 (\omega, \bm{q})$ exhibits
a second jump as a function of frequency. 
For tetragonal symmetry, similar results have been reported
by Norman in Ref.~\onlinecite{norman01}.
To illustrate the fact that the threshold frequencies correspond to 
(local) minima in the two-particle energy, 
we show in Fig.~\ref{Fig: contour} 
the $\bm{k}$ dependence of $E^{\ }_2(\bm{q}, \bm{k})$ 
at the wave vectors $\bm{q}=(1.25 \pi, \pi)$ and $\bm{q}=(\pi, 1.25 \pi)$ 
together with the  
associated scattering vetors between points on the FS  
[Fig.~\ref{Fig: contour}  (c) and Fig.~\ref{Fig: contour} (d)].
In order to isolate the effect of an orthorhombic FS from the effect of a 
subdominant $s$-wave
component we plot in Fig.~\ref{Fig: sepdisp} the dispersion of the threshold 
frequencies for 
$\Delta^{\ }_s=0$, $\delta^{\ }_0 = -0.03$, and $\Delta^{\ }_s=3$~meV, $\delta^{\ }_0 = 0$, 
respectively.

In the case of an orthorhombic FS  and for $\bm{q}$ along the $q^{\ }_x$ direction 
the first scattering process 
[label (1) in Fig.~\ref{Fig: contour}] 
connects points that are further away from the gap \textit{nodes}
(see Fig.~\ref{Fig: fermi surface})
than the corresponding points for $\bm{q}$ along the  $q^{\ }_y$ direction. 
To the contrary, when $\bm{q}$ is on the horizontal axis passing through $(\pi, \pi)$
the scattering vector of the second scattering process
[label (2) in Fig.~\ref{Fig: contour}]
connects points that are further
away from the \textit{antinodes} than the analogous points for $\bm{q}$ 
on the vertical axis.\cite{footnote2}
This behavior reflects itself in the shape of the threshold lines  
[Figs.~\ref{Fig: sepdisp}(a),~\ref{Fig: sepdisp}(b), and~\ref{Fig: fulldisp}]. 
It is found that 
$\omega^{\ }_1(q,\pi)>\omega^{\ }_1(\pi,q)$, 
whereas 
$\omega^{\ }_2(q,\pi)<\omega^{\ }_2(\pi,q)$ for any $q$. 
Since the local minima corresponding to the second scattering process 
are absent in the range $( 0.65 \pi  \lesssim q^{\ }_x \lesssim 0.85 \pi, q^{\ }_y= \pi)$
and $( 1.15 \pi \lesssim q^{\ }_x \lesssim 1.35 \pi, q^{\ }_y= \pi)$ 
the line $\omega^{\ }_2(\bm{q})$ along the $q^{\ }_x$ direction has a gap 
in this momentum range
[Figs.~\ref{Fig: contour}(a) and~\ref{Fig: fulldisp}]. As seen from
Fig.~\ref{Fig: fermi surface}(c) 
the inclusion of a subdominant $s$-wave component 
$\Delta^{\ }_s=3$~meV tilts the vector connecting the SC nodes from 
the diagonal line towards the $x$ direction. 
Hence, the mismatch between the node-to-node vector and
a wave vector $\bm{q}$ along the $q^{\ }_y$ direction is smaller 
than between the node-to-node vector and
a wave vector $\bm{q}$ along the $q^{\ }_x$ direction 
[see Figs.~\ref{Fig: contour}(c) and~\ref{Fig: contour}(d)]. This leads
to a smaller minimum of the particle-hole continuum along the $q^{\ }_y$ direction 
than along the $q^{\ }_x$ direction 
[Figs.~\ref{Fig: sepdisp}(c),~\ref{Fig: sepdisp}(d), and~\ref{Fig: fulldisp}].
Finally, we note that the energy dispersion around the global mimina in 
$E^{\ }_2(\bm{q}, \bm{k})$
for $\bm{q}$ on the horizontal line is flatter than the dispersion for 
$\bm{q}$ on the vertical line  (Fig.~\ref{Fig: contour}), 
which results in a larger jump in $\chi^{\prime\prime}_0(\omega, \bm{q})$ 
in the $q^{\ }_x$ direction than in the $q^{\ }_y$ direction.

As mentioned above the dispersion of the spin excitations tracks the 
behavior of the border to the particle-hole continuum $\omega^{\ }_1(\bm{q})$. 
In Fig.~\ref{Fig: fulldisp}
the position of the resonance peak are represented by 
open diamonds. We find that the downward 
parabola  of the incommensurate peaks has a larger opening angle for 
$\bm{q}$ along the $q^{\ }_x$ direction than for $\bm{q}$ along the $q^{\ }_y$ direction.
The dispersion is flatter in the $q^{\ }_x$ direction leading to incommensurate peaks
that are broader in momentum space for a momentum transfer $\bm{q}$ on the
horizontal axis than for $\bm{q}$ on the vertical axis. Moreover, if
constant energy scans are taken, the incommensurate peaks along the
$q^{\ }_x$ direction are about twice as intense than
those along the $q^{\ }_y$ direction.
This is due to the flatter energy dispersion
of $E^{\ }_2(\bm{q}, \bm{k})$ for $\bm{q}$ on the $x$ axis, and
is in agreement with INS experiments
recently performed by Hinkov \textit{et al.} in 
$\mbox{YBa}^{\ }_2\mbox{Cu}^{\ }_3\mbox{O}^{\ }_{6.85}$, i.e.,
near optimal doping 
(see Fig. 1 in Ref. \onlinecite {hinkov04}). 
The fact that the magnetic response is larger
along the $q_x$ direction compared to the $q_y$ direction for the energy range 
$30$~meV $\leq \omega < 43$ meV is robust as long as the anisotropy in the 
hopping parameters ($t_x < t_y$) dominates
the anisotropy in the SC gap. For example, we have computed 
Fig.~\ref{Fig: im chi q map with full orthorhombic} with the  
band structure of Fig.~\ref{Fig: fermi surface}(b), and found very similar results.
The full
parabolic dispersion of the resonance peak both along the
$q^{\ }_x$- and $q^{\ }_y$ direction still needs to be measured.
For energies smaller than 25meV we find that the presence 
of a subdominant $s$-wave component 
in the SC gap shifts the intensity maxima in 
$\chi^{\prime\prime}_{\hbox{\tiny RPA}}(\omega, \bm{q})$ at 
a constant transfer energy
from the horizontal axis passing through $(\pi, \pi)$ to the vertical axis.

\section{Summary}
\label{sec: summary} 

In summary, we have determined the effect of anisotropic hopping matrix
elements and a mixing of $d$- and $s$-wave symmetry of the gap on
the dynamical magnetic susceptibility of \mbox{high-$T^{\ }_c$}
cuprates within a Fermi-liquid-based theory. For transfer energies
smaller than the resonance energy, $\omega^{\ }_{res}$, we find
strongly anisotropic spin excitations on the horizontal and
vertical axes of the magnetic BZ. The inclusion of anisotropic
hopping parameters leads to a distortion of the square-like
excitation pattern at a constant transfer energy to a rhombus shape. 
For $t^{\ }_x < t^{\ }_y$ and within
the energy window $1/2 \, \omega^{\ }_{res} < \omega < \omega^{\ }_{res}$ we have
shown that the spin excitations along the $q^{\ }_x$ direction are about
twice as intense
than the ones along the $q^{\ }_y$ direction.
Furthermore, we predict considerable 
differences in the dispersion of the resonance peak along the 
$(q^{\ }_x,\pi)$- and $(\pi, q^{\ }_y)$-axes, 
respectively (see Fig.~\ref{Fig: fulldisp}). The peaks
along the $q^{\ }_x$ direction are both further apart and broader in momentum space
compared to the peaks along the $q^{\ }_y$ direction.

The effect of a subdominant $s$-wave component in the
superconducting gap is most prominent at small energies of about
$\simeq 1/2 \, \omega^{\ }_{res}$. Assuming $\Delta^{\ }_{s} >0$, as 
demanded by ARPES
measurements,~\cite{lu01} the subdominant $s$-wave component
results in a rotation of the intensity maxima by 90$^o$ relative
to the excitation pattern at energies $1/2\, \omega^{\ }_{res} \lesssim
\omega < \omega^{\ }_{res}$, and the spin gap becomes strongly
anisotropic. 

Between the resonance energy and a transfer energy of up to 50$\%$ larger than
the resonance energy, the spin response remains
anisotropic  with a suppression of the intensity along the
$q^{\ }_x$ direction. The anisotropy between the spin response
along the inequivalent directions $q^{\ }_x$ and  $q^{\ }_y$
decreases with an increasing transfer energy above the  resonance energy.
Intensities are negligible at transfer energies 400$\%$ 
larger than the resonance energy in sharp contrast to what is measured for
$\mbox{La}^{\ }_{15/8}\mbox{Ba}^{\ }_{1/8}\mbox{Cu}\mbox{O}^{\ }_{4}$
in Ref.~\onlinecite{tranquada04}.

\acknowledgments
It is our pleasure to thank H.~Yamase, S.~Pailhes, V.~Hinkov, B.~Keimer, J.~Mesot, and W.~Metzner 
for useful discussions and H.~Hilgenkamp for the prepublication copy of
Ref.~\onlinecite{smilde05}.
This work was supported by the Swiss National Science Foundation
under Grant No. 200021-101765/1 and the NCCR MaNEP. D.~M. acknowledges financial
support from the Alexander von Humboldt foundation.


\begin{thebibliography}{99}
\bibitem{footnote}
%
We would like to emphasize that our single-band model
is a phenomenological one. It should be thought of
as capturing the low energy physics of 
a more realistic microscopic model made of several bands
describing, for example, the bilayer structure and the CuO chains in the
YBCO family.
Assuming orthorhombicity and the crystal structure of YBCO there are two competing 
phenomena. In an effective one-band Hubbard model it is natural to account for a lattice spacing 
along the $b$ direction that is larger than along the $a$ direction by hopping amplitudes 
$t^{\ }_x > t^{\ }_y$ . 
However, the presence of the chain allows for a higher mobility for a charge carrier along the 
chain, which would lead to the opposite effect (cf.~Ref.~\onlinecite{Li04}).
What wins depends on microscopic details 
that are beyond our scope. Due to the much stronger anisotropy caused
by the chains, one might naively expect that $t^{\ }_y > t^{\ }_x$, which is
consistent with INS data.
\bibitem{footnote2}
In Fig.~\ref{Fig: contour}(c) the scattering vector of type (2) is not shown, since the
corresponding minima in $E^{\ }_2(\bm{q}, \bm{k})$ do not exist.
\end{thebibliography}
\end{document}